\documentclass[aps, prd, twocolumn, superscriptaddress, nofootinbib, floatfix, noshowpacs]{revtex4-2}
\pdfoutput=1
\newif\ifcomment
\usepackage{amsmath,amssymb}
\usepackage{color,url}
\usepackage{listings}
\usepackage{slashed}
\usepackage[pdftex]{graphicx}
\usepackage{epstopdf}
\usepackage{epsfig}
\usepackage{grffile}
\usepackage{relsize}
\usepackage{float}
\graphicspath{{./img/}}
\usepackage{soul}

\usepackage{CJKutf8}

\usepackage[mathscr,scaled=1.15]{urwchancal}
\DeclareFontFamily{OT1}{pzc}{}
\DeclareFontShape{OT1}{pzc}{m}{it}%
{<-> s * [1.15] pzcmi7t}{}
\DeclareMathAlphabet{\mathpzc}{OT1}{pzc}{m}{it}

\definecolor{purple}{rgb}{0.5,0,0.5}
\definecolor{blue}{rgb}{0.0,0,0.9}
\definecolor{prdblue}{rgb}{0.133,0.118,0.498}
\usepackage[colorlinks=true, pdfstartview=FitV, linkcolor=prdblue, citecolor= prdblue, urlcolor=prdblue]{hyperref}

\RequirePackage{comment}

\newcommand{\beq}{\begin{equation}}
\newcommand{\eeq}{\end{equation}}
\newcommand{\ba}{\begin{array}}
\newcommand{\ea}{\end{array}}
\newcommand{\bea}{\begin{align}}
\newcommand{\eea}{\end{align}}
\newcommand{\bi}{\begin{itemize}}
\newcommand{\ei}{\end{itemize}}
\newcommand{\ben}{\begin{enumerate}}
\newcommand{\een}{\end{enumerate}}
\newcommand{\bc}{\begin{center}}
\newcommand{\ec}{\end{center}}
\newcommand{\bl}{\begin{flushleft}}
\newcommand{\el}{\end{flushleft}}
\newcommand{\br}{\begin{flushright}}
\newcommand{\er}{\end{flushright}}

\begin{document}
\begin{CJK*}{UTF8}{gbsn}

\title{$\,$\\[-6ex]\hspace*{\fill}{\normalsize{\sf\emph{Preprint no}.\
NJU-INP 121-26}}\\[1ex]
Quest for an Understanding of Pion and Kaon Structure}

\author{Zhen-Ni Xu (徐珍妮)%
       $^{\href{https://orcid.org/0000-0002-9104-9680}{\textcolor[rgb]{0.00,1.00,0.00}{\sf ID}}}$}
\email[]{zhenni.xu@dci.uhu.es}
\affiliation{Department of Integrated Sciences and Center for Advanced Studies in Physics, Mathematics and Computation, \href{https://ror.org/03a1kt624}{University of Huelva}, E-21071 Huelva, Spain}

\author{Daniele Binosi%
    $^{\href{https://orcid.org/0000-0003-1742-4689}{\textcolor[rgb]{0.00,1.00,0.00}{\sf ID}}}$}
\affiliation{European Centre for Theoretical Studies in Nuclear Physics
            and Related Areas  (\href{https://ror.org/01gzye136}{ECT*}), Villa Tambosi, Strada delle Tabarelle 286, I-38123 Villazzano (TN), Italy}

\author{Craig D. Roberts%
       $^{\href{https://orcid.org/0000-0002-2937-1361}{\textcolor[rgb]{0.00,1.00,0.00}{\sf ID}}}$}
\email[]{cdroberts@nju.edu.cn}
\affiliation{School of Physics, \href{https://ror.org/01rxvg760}{Nanjing University}, Nanjing, Jiangsu 210093, China}
\affiliation{Institute for Nonperturbative Physics, \href{https://ror.org/01rxvg760}{Nanjing University}, Nanjing, Jiangsu 210093, China}

\author{Jos\'e Rodr\'iguez-Quintero%
       $^{\href{https://orcid.org/0000-0002-1651-5717}{\textcolor[rgb]{0.00,1.00,0.00}{\sf ID}}}$}
\affiliation{Department of Integrated Sciences and Center for Advanced Studies in Physics, Mathematics and Computation, \href{https://ror.org/03a1kt624}{University of Huelva}, E-21071 Huelva, Spain}

\date{
2026 August 08
}

\begin{abstract}
The emergence of massless (Nambu-Goldstone) bosons in association with a dynamically global broken symmetry is a long known and widespread phenomenon in physics. However, practically nothing is known about the expressions of Nambu--Goldstone boson character on the internal structure of these bound states. Indeed, their structure is often ignored. In strong interactions, pions and kaons are the (would-be) Nambu-Goldstone bosons and experiments underway or planned at existing or anticipated high-energy, high-luminosity facilities will gather data that it is hoped will enable maps to be drawn of their internal structure. Meanwhile, theory and phenomenology find themselves in something of a quagmire. Herein, we provide a snapshot of the current status, highlighting issues under debate and identifying areas that deserve greater attention so that best use can be made of what is likely to be a huge volume of data delivered in the next decade or so.
\\[1ex]
\end{abstract}

\maketitle

\end{CJK*}

\noindent\textbf{Introduction}\,---\,%
In what is today known as the standard model of particle physics (SM), strong interactions are described by quantum chromodynamics (QCD) \cite{Fritzsch:1973pi, Pickering:1984tk, Politzer:2005kc, Wilczek:2005az, Gross:2005kv}.
This is a Poincar\'e-invariant quantum gauge field theory whose one-line Lagrangian is expressed in terms of gluon and quark parton fields \cite{Fritzsch:1973pi, Pickering:1984tk}.
The QCD dynamical charge is called ``colour'' and both gluons and quarks are colour charged; so, unlike the electric charge neutral photons of quantum electrodynamics, gluon gauge bosons self-interact at leading order in perturbation theory.
These gluon self interactions have long been argued to provide the foundation for gluon and quark ``confinement''.
Namely, whereas electrons and photons are readily isolated in experiments, gluon and quark partons are locked within colour-neutral hadrons, \emph{viz}.\ neutrons, protons, pions, kaons, etc.
Neither gluons nor quarks have ever been detected in isolation; instead, they are trapped within  fm-sized ($10^{-15}\,$m radius) hadron bound states.

Anyone who has ever considered the periodic table of elements will feel comfortable with neutrons and protons (nucleons) -- they are the building blocks of nuclei.  Notwithstanding this degree of higher-level familiarity, much remains unknown about their internal QCD-determined structure.
This is despite sixty-years of high-energy experiments, \emph{e.g}., deep inelastic scattering and Drell-Yan processes \cite{Roberts:1990ww, Ellis:1991qj}, that have probed into the heart of nucleons and delivered huge amounts of data.
The problem lies in the lack of a bridge between those data and QCD itself.
Essentially nonperturbative tools are necessary to complete that bridge and only today are robust methods of this type beginning to be deployed.

Given this as the state-of-the-art for protons, which can readily be prepared in stable target configurations, it should not be surprising that very little is known about the internal structure of mesons like pions, $\pi$, and kaons, $K$, whose lifetimes are $\sim 10^{-8}\,$s.
In these cases, stable targets are impossible to construct.
Instead, one must be imaginative with experiments at collider facilities; see, \emph{e.g}., Refs.\,\cite[Sec.\,9]{Roberts:2021nhw}, \cite{Arrington:2021biu, Chavez:2021koz, Quintans:2022utc, Lu:2025bnm, Chang:2024rbs, Chang:2025ind}.
Even with this approach, kaon data will always remain relatively scarce because kaons are hard to produce.
All these and related difficulties are nevertheless worth overcoming because pions and kaons are special, not least because they are Nature's most fundamental (near) Nambu-Goldstone bosons.

In the SM, the one phenomenologically understood source of mass for elementary particles is the Higgs boson \cite{Englert:2014zpa, Higgs:2014aqa}, \emph{e.g}.,
Higgs boson couplings into QCD give quark partons their Lagrangian (current) masses.
If those couplings are eliminated, then one implements the ``chiral limit'' and QCD is a mass/length-scale invariant theory.
Classically, scale invariant theories cannot support any bound states; thus, the fundamental entities necessary for the construction and stabilisation of atomic nuclei (protons, neutrons, pions, kaons, etc.) cannot exist \cite{Roberts:2016vyn}.

Restoring Higgs boson couplings, then for the up ($u$) and down ($d$) quark partons, one obtains light current masses $\hat m_d \approx 2 \hat m_u\approx 10\,m_e$, where $m_e$ is the electron mass \cite[PDG]{ParticleDataGroup:2024cfk}.
In this case, the pion mass is $m_\pi \approx 300\,m_e \approx m_\tau$, \emph{i.e}., the $\pi$-meson and $\tau$-lepton are roughly degenerate, whereas the proton mass is far greater, $m_p\approx 2\,000\,m_e$.
Both pions, which are largely responsible for binding atomic nuclei \cite{Yukawa:1935xg}, and protons are built from valence light quark and/or antiquark degrees of freedom (dof); so, the following question immediately begs for an answer: How does a nearly scale invariant theory -- characterised by electron-size masses -- produce a $\pi$-meson with a $\tau$-lepton-like mass and, simultaneously, a proton with a nuclear-size mass?
In addressing this puzzle, one steps immediately into the realm of emergent strong interaction phenomena (ESIP) and, particularly, the enigma  of emergent hadron mass (EHM) \cite{Ding:2022ows, Achenbach:2025kfx, Binosi:2026tre}.

\begin{figure}
\centerline{\includegraphics[clip, width=0.3\textwidth]{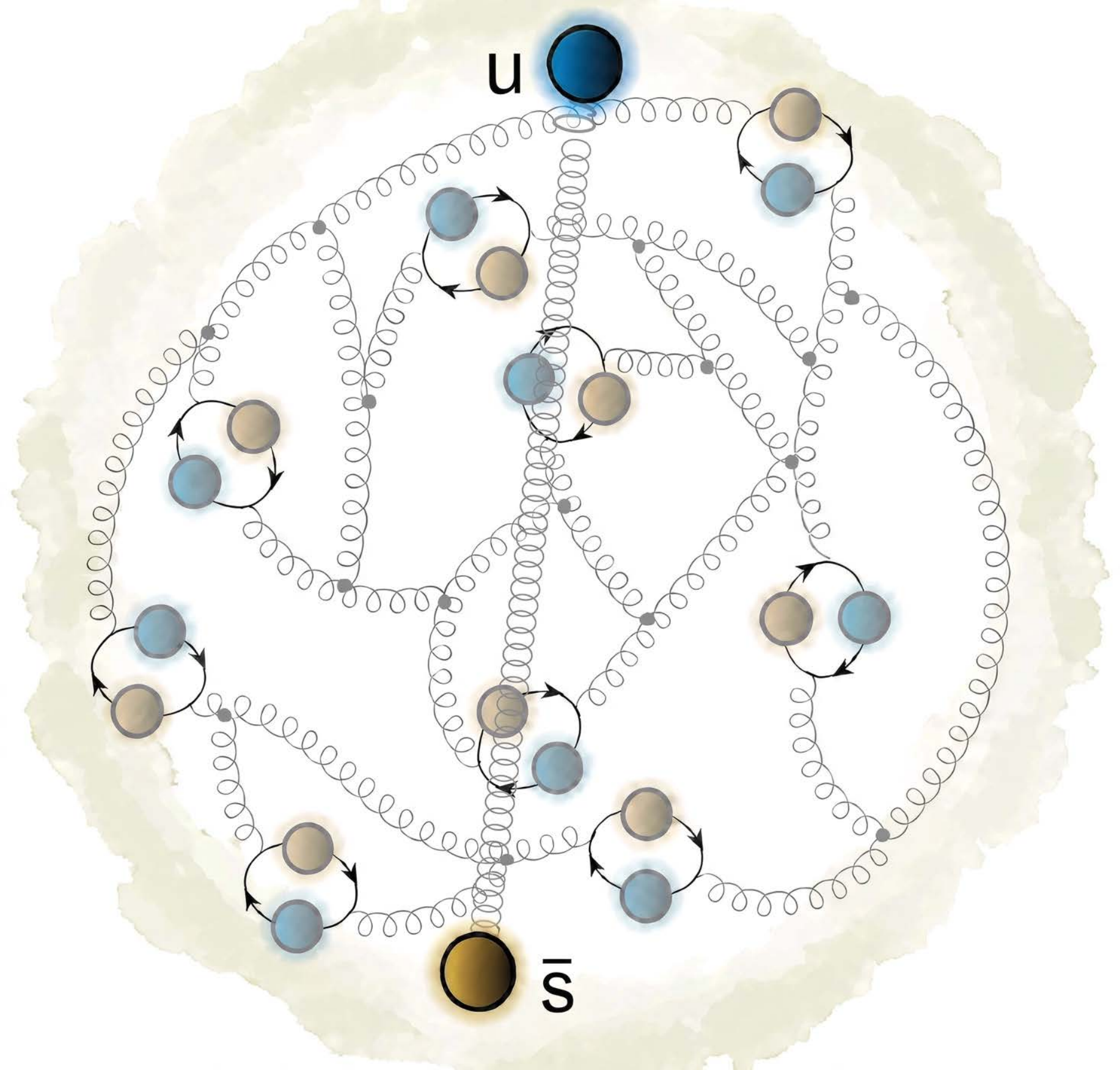}}

\vspace*{-2ex}

\caption{\label{F1kaon}
The kaon, $K^+$, contains one valence $u$-quark, one valence $\bar s$-quark, and, in QCD, infinitely many gluons and sea quarks.
In terms of valence quarks, $K^- = \bar u s$, $K^0 = d\bar s$, $\bar K^0 = \bar d s$.  Insofar as strong interactions are concerned, all four states are equivalent.
}
\end{figure}

Kaons are constituted from one valence light quark/antiquark and one valence strange ($s$) antiquark/quark: an impression of $K^+$ is depicted in Fig.\,\ref{F1kaon}.  The strange quark current mass is $\hat m_s \approx 27 (\hat m_u + \hat m_d)/2$; so, much larger than that of the $u$, $d$.
Yet, the $s$-quark is not truly heavy.
Moreover, $m_K \approx 3.6\, (\hat m_u + \hat m_s) $, which may be compared with $m_\pi \approx 10 (\hat m_u + \hat m_d) $ \cite{Roberts:2020udq}.
Evidently, that part of the kaon mass generated by the Higgs is significantly larger than that of the pion \cite[Sec.\,2]{Ding:2022ows}.
Notwithstanding, pions and kaons are all massless in the chiral limit, \emph{i.e}., they are Nambu-Goldstone bosons.
From the QCD perspective, very precise symmetry-driven cancellations take place between interactions inside these hadrons to ensure that, with no fine tuning, they are massless when the quark current masses vanish \cite[Sec.\,3]{Roberts:2016vyn}.

Against this background one can understand that pions and kaons, and comparisons between their properties, provide clear (actually the cleanest \cite{Roberts:2016vyn}) windows onto emergent features of hadron mass and structure and the modulation of these features by Higgs boson couplings into QCD.  This realisation has been the ignition key for both experimental $\pi$, $K$ structure programmes worldwide \cite{Arrington:2021biu, Chavez:2021koz, Quintans:2022utc, Lu:2025bnm, Chang:2024rbs, Chang:2025ind} and a rapid expansion in theory efforts to predict and understand their composition \cite{Roberts:2021nhw}.
The goal of these activities is to draw maps of the distributions of gluons and quarks within these especially interesting strong interaction bound states:
are there differences between in-pion and in-kaon distributions;
what are they;
why do they appear;
and how do in-pion and in-kaon distributions differ from those in the proton?
Answers to these questions are essential to developing an explanation for the origin of mass.

\smallskip

\noindent\textbf{Parton Distribution Functions from Data}\,---\,%
In Poincar\'e-in\-va\-riant quantum field theory and, therefore, QCD, parton number is not conserved by interactions.  Amongst other things, this entails that whilst a given hadron is \emph{defined} by its valence dof content, which is conserved, \emph{e.g}., $u\bar s$ for the $K^+$, at any measurement scale whereat the quantities of interest can be extracted, these valence quark partons are accompanied by infinitely many glue and sea quark partons  (``sea'' means quark + antiquark parton pairs); see Fig.\,\ref{F1kaon}.  So, the charts sought are properly expressed in terms of parton number densities, which are typically called parton distribution functions (DFs) \cite[Ch.\,4]{Ellis:1991qj}.

In cases that involve spin/helicity-averaged cross-sections, these DFs are defined in terms of the modulus-squared of the wave function of the hadron involved \cite{Diehl:2000xz}; hence, they are non-negative functions of their arguments.
To specify those arguments, it is necessary to state that there is only one approach to field theory quantisation which delivers wave functions that possess the sort of probability density interpretation associated with Schr\"odinger wave functions in nonrelativistic quantum mechanics, \emph{viz}.\ the light-front formulation \cite{Brodsky:1997de, Brodsky:2022fqy}.
In order to avoid unnecessary complication herein, it is worth specifying at the outset that we will only consider the so-called collinear DFs.  These functions depend on just one variable, \emph{i.e}., $x$, the light-front fraction of the target hadron's four-momentum carried by the struck parton.

An intuitive picture of DF physics is readily obtained by imagining that one is analysing scattering in the infinite momentum frame  \cite[Ch.\,4]{Ellis:1991qj}, wherein the four-momentum of a target hadron $H$, with mass $m_H$, is $p=(0,0,P,iP)$, where $P \gg m_H$, and the momentum of a struck in-$H$ parton is $k=x p$.  All momenta transverse to the direction of motion are negligible in the infinite momentum frame.
(We work with a Euclidean formulation of QCD.  Described, \emph{e.g}., in Ref.\,\cite[Sec.\,1]{Ding:2022ows}, this convention will not have any impact on our discussion.)

As noted above, parton DFs are extracted from data obtained in high-energy scattering experiments.  High energy is necessary because one must work on a kinematic domain whereupon the associated cross section can reliably be approximated as a convolution.  Namely, some structure function characteristic of the scattering process can be written in the following (or a qualitatively analogous) form:
\begin{equation}
F^H(x;\zeta) = \int_{x}^1 (dy /y) \sigma_{pRM}(x/y;\zeta) {\mathpzc f}^H(y;\zeta)\,,
\label{factorisation}
\end{equation}
where the energy scale $\zeta \gg m_p$;
$\sigma_{pRM}(z;\zeta)$ is a good approximation to the (supposedly target-independent) parton level scattering cross-section (hard scattering kernel) at $\zeta$;
and ${\mathpzc f}^H(y;\zeta)$ is an in-hadron parton DF at this scale.
(Here, ${\mathpzc f} = \mathpzc q/\bar {\mathpzc q}$, ranging over flavours, are quark/antiquark DFs and ${\mathpzc f} = \mathpzc g$ is the glue DF.)
The DF is the number-density of partons in $H$ that carry a light-front momentum-fraction in the neighbourhood of $x$.

When discussing DFs, it is common to work with their Mellin moments:
\begin{equation}
\langle x^n \rangle_{{\mathpzc f}^H}^{\zeta}
= \int_0^1 dx\, x^n\, {\mathpzc f}^H(x;\zeta)\,.
\end{equation}
The $n=1$ moment is the mean light-front fraction of the hadron's momentum carried by the given parton species.
For any physical system, the moments of a realistic DF possess the following property:
\begin{equation}
\label{MomOrdering}
\forall n \in \mathbb N_0\,|\, \langle x^{n+1} \rangle_{{\mathpzc f}^H}^{\zeta}
< \langle x^{n} \rangle_{{\mathpzc f}^H}^{\zeta}\,.
\end{equation}
In principle, if all moments of a given DF are known, then the pointwise form of the object function is recoverable.  On the other hand, if only a few moments are known (low-order in typical cases), then the $x$-dependence of any reconstructed DF is subjective, being determined by practitioner-dependent fitting choices/reconstruction algorithms.

Two significant impediments are encountered when attempting to use Eq.\,\eqref{factorisation}.
\emph{(i}) $F$ is a collection of data, $\sigma_{pRM}$ is a calculated quantity; so to obtain ${\mathpzc f}^H$, one must solve an ``inverse problem''.
Solving such (mathematically ill-posed) problems is notoriously difficult because the solution is typically not unique and noise in the data or uncertainty in the kernel ($\sigma_{pRM}$) can lead to very different results.  (Indeed, domains of negative support can be obtained in functions that are known, by definition, to be non-negative \cite{Candido:2023ujx}.)
Consequently, the solution obtained depends on the algorithms employed.
(\emph{ii}) No matter how precise the data may be, the pointwise behaviour ($x$ dependence) of whatever answer is obtained depends critically on the form of $\sigma_{pRM}$, \emph{i.e}., the parton reaction model assumed to be valid at the scale $\zeta$.  Today, in QCD, the character of a good approximation for $\sigma_{pRM}$ is much debated; see, \emph{e.g}., Ref.\,\cite{Cui:2021mom} for a discussion of the pion case.  The model chosen for $\sigma_{pRM}$ is a subjective (practitioner-dependent) ingredient in all existing fits.

Two other issues, which cannot readily be overcome in fitting approaches, are that practitioners must define a starting scale whereat some functional forms/representations of the valence quark, sea quark, and glue DFs are specified and from which perturbative DGLAP evolution can be implemented.  Moreover, that the assumed DFs are independent, \emph{viz}.\ there are no connections/correlations between the DFs of various parton species.
QCD, if an internally consistent theory, would deliver strongly correlated DFs for participating parton species at all scales and it is improbable that any practitioner-dependent choice made at some arbitrary resolving scale would capture such dynamically induced relationships.

An array of collaborations have attempted to extract pion DFs from available data: recent fit results \cite{Novikov:2020snp, Pasquini:2023aaf, Kotz:2025lio, Barry:2025wjx} are available at the Les Houches accord parton distribution function repository (LHAPDF) \cite{Whalley:2005nh}.
As indicated above, any phenomenological extraction of $K$ DFs is subject to large uncertainties because available kaon data are scarce \cite[NA3K]{Badier:1980jq}.
Notably, therefore, only Ref.\,\cite[lJAM25]{Barry:2025wjx} reports an attempt at the simultaneous inference of $\pi$, $K$ DFs.
However, owing to the scarcity of data, lJAM25 introduced additional subjectivity into the analysis by choosing to constrain the fitting procedure via selected inputs from a single numerical simulation of lattice-regularised QCD (lQCD) \cite{Alexandrou:2021mmi}.
Notwithstanding, and despite other acknowledged limitations (\emph{e.g}., if reliable, then on $x>0.1$), only the lJAM25 fits are considered hereafter because we focus on unified phenomenological and theoretical treatments of $\pi$, $K$ DFs. 

\smallskip

\noindent\textbf{Effective Charge, All-Orders Evolution, and Empirical DFs}\,---\,%
Equation~\eqref{factorisation} highlights that DFs depend on the measurement scale.
However, this does not mean that measurements and analyses must be performed anew at each desired scale because the DGLAP evolution equations \cite{Dokshitzer:1977sg, Gribov:1971zn, Lipatov:1974qm, Altarelli:1977zs} provide the QCD connection between DFs at any two scales $\zeta, \zeta^\prime \gtrsim 2 m_p$.

One can define a nonperturbative extension of DGLAP evolution by using the theory of QCD effective charges, introduced in
Refs.\,\cite{Grunberg:1980ja, Grunberg:1982fw} and reviewed in Ref.\,\cite{Deur:2023dzc}.
This theory enables one to implicitly identify an effective charge (running coupling), $\alpha_{1\ell}(k^2)$, which, when used to integrate the leading-order perturbative DGLAP equations, defines an evolution scheme for \emph{all} parton DFs that is both all-orders exact and applicable at all scales \cite{Yin:2023dbw}.
Significantly, the pointwise form of $\alpha_{1\ell}(k^2)$ is largely irrelevant; for most purposes, one need only know that such a charge exists.  Nevertheless, the QCD process-independent (PI) running coupling calculated in Ref.\,\cite{Cui:2019dwv} has all required properties.
Given such an effective charge, then it follows that there is a scale, $\zeta_{\cal H}$, the same for \emph{all} hadrons, whereat \emph{all} properties of a given hadron are carried by its valence dof.
At $\zeta_{\cal H}$, DFs associated with glue and sea quarks are zero.

This effective charge scheme is called AO evolution.
It is the universality of the approach -- all DFs, all scales, and all hadrons -- that imbues the AO scheme with its appeal and power.

Like the pointwise form of $\alpha_{1\ell}(k^2)$, the numerical value of $\zeta_{\cal H}$ is immaterial.
If one chooses to use the PI coupling, then \cite{Cui:2021mom}:
$\zeta_{\cal H} = 0.331(2)\,{\rm GeV}$.
A consistent value, $\zeta_{\cal H} = 0.350(44)\,{\rm GeV}$, is obtained via analysis of results relating to the pion valence quark DF that were obtained using numerical simulations of lQCD \cite{Lu:2023yna}.

As explained above, using the AO approach and assuming ${\cal G}$-parity symmetry \cite{Lee:1956sw}, which is a good approximation in Nature, the $\pi^+$ and $K^+$ valence quark/antiquark DFs possess the following properties:
\begin{subequations}
\label{DFsymmetry}
\begin{align}
{\mathpzc u}_V^\pi(x;\zeta_{\cal H}) & = \bar {\mathpzc d}_V^\pi(1-x;\zeta_{\cal H})
= {\mathpzc u}_V^\pi(1-x;\zeta_{\cal H}) \,,\\
{\mathpzc u}_V^K(x;\zeta_{\cal H}) & = \bar {\mathpzc s}_V^K(1-x;\zeta_{\cal H})\,.
\end{align}
\end{subequations}
Consequently, the hadron-scale $\pi^+$ and $K^+$ momentum fractions are:
\begin{equation}
\langle x \rangle_{\bar{\mathpzc d}_V^\pi}^{\zeta_{\cal H}}= \langle x \rangle_{{\mathpzc u}_V^\pi}^{\zeta_{\cal H}}  = \tfrac{1}{2}\,,
\quad
\langle x \rangle_{{\mathpzc u}_V^K}^{\zeta_{\cal H}}
+\langle x \rangle_{\bar{\mathpzc s}_V^K}^{\zeta_{\cal H}}  = 1\,.
\label{momfracs}
\end{equation}
Charge conjugation gives the results for $\pi^-$, $K^-$.

Working with kaon \cite[NA3K]{Badier:1980jq} and pion \cite[E615]{Conway:1989fs} Drell-Yan data, and the analysis of the latter described in Refs.\,\cite{Aicher:2010cb, Chang:2014lva}, Ref.\,\cite{Xu:2024nzp} used structure-function-constrained probability-weighted ensembles of valence DF replicas and AO evolution to simultaneously obtain pointwise profiles for all kaon and pion DFs without reference to theories of hadron structure.  In the following, we will denote the Ref.\,\cite{Xu:2024nzp} results as ``empirical''.

\smallskip

\begin{figure*}
\begin{tabular}{ccc}
{\large\sf A} \hspace*{26em} & $\;$ & {\large \sf B} \hspace*{26em} \\[-4ex]
\includegraphics[clip, width=0.46\textwidth]{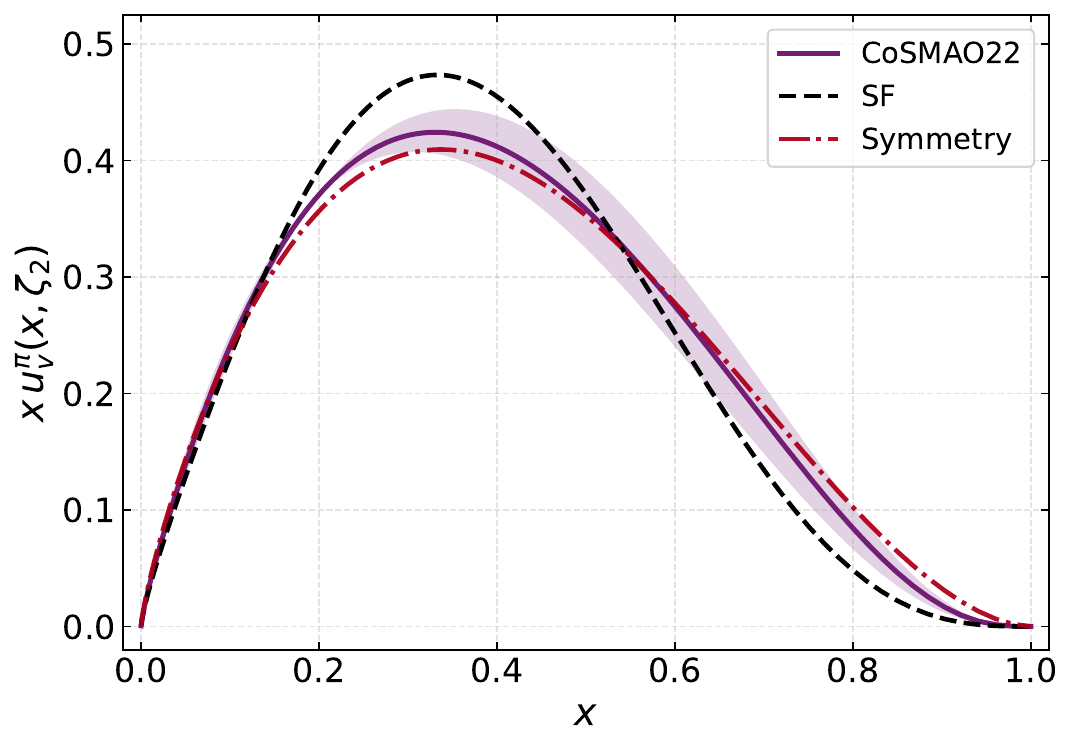} & $\;$ &
\includegraphics[clip, width=0.46\textwidth]{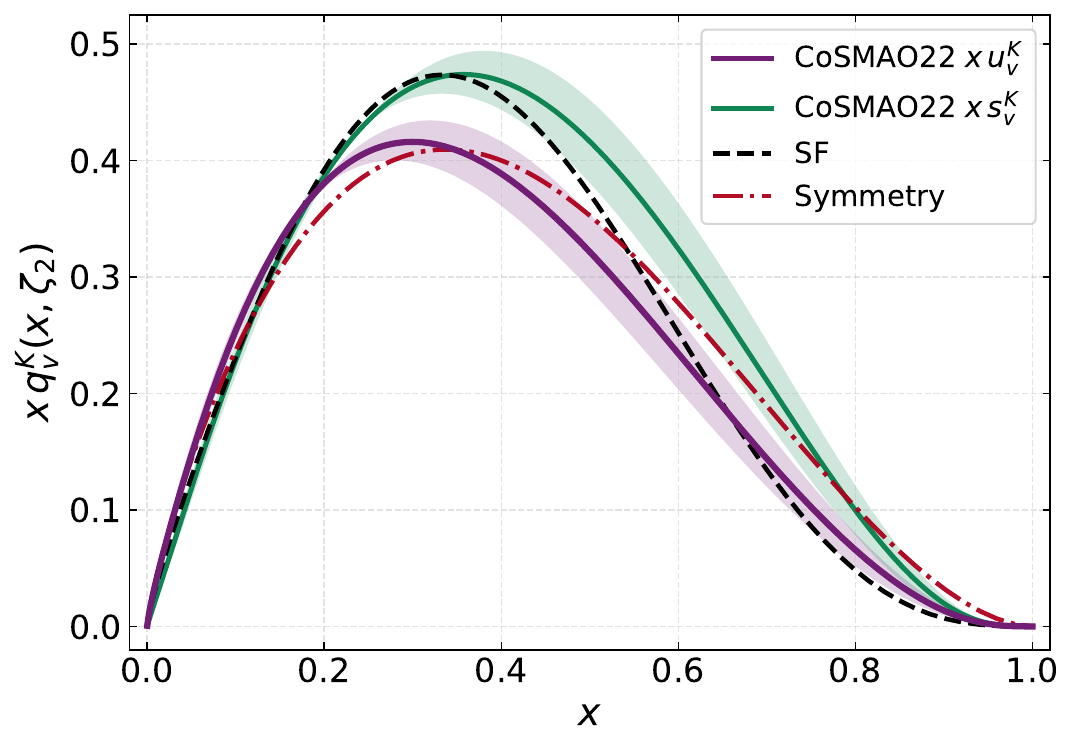}
\end{tabular}


\vspace*{-2ex}

\caption{\label{FpiKDF}
CSM (CoSMAO) predictions \cite{Cui:2020tdf} for pointwise behaviour of pion ({\sf A}) and kaon ({\sf B}) valence degree-of-freedom DFs at $\zeta=\zeta_2 := 2\,$GeV.
The comparison curves are the no-ESIP (SF) result, associated with Eq.\,\eqref{sfDF}, and the model-independent ESIP benchmark prediction (Symmetry) \cite{Wang:2025usl}.
}
\end{figure*}

\noindent\textbf{Continuum DF Predictions}\,---\,%
Continuum Schwinger function methods (CSMs), \emph{e.g}., Dyson-Schwinger equations \cite{Ding:2022ows} and/or the functional renormalisation group \cite{Dupuis:2020fhh}, provide a systematic, symmetry preserving approach to solving QCD.  Importantly, they are applicable at all momentum scales, \emph{i.e}., on both the perturbative and nonperturbative domains.
Some highlights of their recent applications to pion and kaon observables may be found in Refs.\,\cite{Roberts:2021nhw, Ding:2022ows, Raya:2024ejx, Yao:2024drm, Xing:2025eip, Xiao:2025cqz}.
Of particular significance herein are recent CSM applications to prediction of the \emph{pointwise} behaviour of $\pi$, $K$ DFs \cite{Cui:2020tdf}.  These predictions are also available at the LHAPDF repository, under the ``CoSMAO22'' label.

Notably, CSMs veraciously describe the (near) Nambu-Goldstone boson character of pions and kaons \cite{Maris:1997tm}, including the large relative momentum behaviour of their Poincar\'e-covariant bound-state wave functions \cite{Politzer:1976tv}.   It is these features that enable CSMs to deliver objective information on EHM and its observable expressions.
Nambu-Goldstone bosons and their wave functions present a challenge for alternative continuum approaches to hadron structure; see, \emph{e.g}., Refs.\,\cite{dePaula:2022pcb, Liu:2023yuj, Forshaw:2024mrh, Lan:2025fia, Puhan:2026ppv}; so, whilst they express novel perspectives, we do not consider them further herein.

As stressed above, a particular feature of CSMs is that, since they expose and employ gluon and quark scattering kernels that emerge as solutions of QCD gauge sector dynamics \cite{Binosi:2014aea, Ferreira:2023fva, Binosi:2026tre}, they deliver predictions which conform with QCD constraints on the endpoint behaviour of parton DFs.
Namely, as expected from Ref.\,\cite{Brodsky:1994kg} and discussed in Ref.\,\cite{Cui:2021mom}:
\begin{equation}
{\mathpzc u}_V^\pi(x;\zeta_{\cal H}) \stackrel{x\simeq 1}{\propto} (1-x)^{\beta = 2}\,,
\label{DFpower}
\end{equation}
with the same results for ${\mathpzc u}_V^K(x;\zeta_{\cal H})$, $\bar {\mathpzc s}_V^K(x;\zeta_{\cal H})$.  Given the properties of DGLAP evolution, it follows that at any scale $\zeta>\zeta_{\cal H}$ for which DF inference from data is possible via Eq.\,\eqref{factorisation}, the large-$x$ power on all $\pi$, $K$ valence distributions is $\beta = 2+ \gamma(\zeta)$, where $\gamma(\zeta)>0$.
(See, \emph{e.g}., Ref.\,\cite[Sec.\,4]{Cui:2020tdf}, which gives a mathematical formula for $\gamma(\zeta)$.)
AO evolution also delivers nonzero glue and sea DFs at any $\zeta>\zeta_{\cal H}$: the large-$x$ power on glue DFs is $\beta_g \approx \beta + 1$ and that on sea DFs is $\beta_S \approx \beta + 2$.

Any phenomenologically obtained DF fit results that do not comply with these constraints are inconsistent with an underlying theory based on vector boson exchange interactions between gluon and quark partons \cite[Sec.\,7]{Cui:2021mom}.
In our view, much could be learnt about QCD reaction models, \emph{i.e}., the form of $\sigma_{pRM}$, by implementing Eq.\,\eqref{DFpower} and its corollaries in fitting algorithms.

When using CSMs, one basic challenge is encountered.
Namely, a quantum field theory is defined by a countable infinity of Schwinger functions, each one labelled by the number, $m$, of external lines, and higher-order functions are related to those of lower order via integral or integro-differential equations.
For instance, the dressed quark two-point Schwinger function (propagator) is determined by an integral equation in whose kernel appears the gluon propagator and gluon-quark three-point function \cite[Fig.\,2.4]{Roberts:1994dr}.
Each of these last two functions satisfies its own integral equation, which involves other three- and four-point functions.
This pattern continues along the increasing $m$-point chain.

Plainly, one cannot simultaneously solve infinitely many equations; so for any given matrix element, a set of truncations must be introduced.
Such truncations are today seen to express reliable approximations for two reasons.
(\emph{i}) QCD symmetries are crucial and modern truncation schemes are built such that those symmetries germane to a given calculation are preserved; see the discussions in Refs.\,\cite{Munczek:1994zz, Bender:1996bb, Bicudo:2001jq, Ding:2019lwe} and citations thereof.
(\emph{ii}) A given $m$-point Schwinger function describes $m$-particle interactions.  With increasing $m$, the likelihood of $m$ particles ``finding'' each other, in order to interact, diminishes rapidly.  Moreover, the process-independent interaction strength, $\hat\alpha(r)$ \cite{Cui:2019dwv}, vanishes with decreasing interparton separation (asymptotic freedom \cite{Politzer:2005kc, Wilczek:2005az, Gross:2005kv}) and is bounded above at long range $\hat\alpha(r\gg 1/m_\pi) \lesssim \pi$.
It follows that, with increasing $m$, the contribution of higher $m$-point functions to any matrix element is rapidly suppressed.

Finally, of course, if any questions remain, then the accuracy of a given truncation can be tested by comparing its results with those obtained using a higher-order improved truncation.
The predictions in Ref.\,\cite{Cui:2020tdf}, available at LHAPDF, are based upon a nonperturbatively-improved truncation that has been validated in this way for $\pi$, $K$ observables: the valence dof DFs are depicted in Fig.\,\ref{FpiKDF} at the scale $\zeta= \zeta_2:=2\,$GeV.
Evidently, CSMs predict that Higgs-boson introduced differences between light and strange quarks are expressed in modest but noticeable strength redistributions between ${\mathpzc u}^K(x)$ and $\bar {\mathpzc s}^K(x)$, with the in-kaon light-front momentum fraction of the $\bar s$ valence quark being roughly 20\% larger than that of the $u$ valence quark.

Two comparison curves are also drawn in Fig.\,\ref{FpiKDF}.  One arrives at the first by recalling that, in discussing meson parton distribution amplitudes, DAs, the known scale-independent/free QCD asymptotic form is $\varphi^{\rm as}(x) = 6 x(1-x)$ \cite{Lepage:1980fj}.
A hadron-scale analogue for DFs is  \cite{Chang:2014lva, Yao:2025xjx}:
\begin{equation}
{\mathpzc q}^{\rm sf}(x;\zeta_{\cal H}) = 30 x^2 (1-x)^2 \propto |\varphi^{\rm as}(x)|^2\,.
\label{sfDF}
\end{equation}
This DF describes the valence dof structure of a pion-like system in the absence of ESIP.  After AO evolution $\zeta_{\cal H} \to \zeta_2$, one obtains the dashed-black curve in Fig.\,\ref{FpiKDF}.
The other comparison curve is a model-independent DF result that emerges simply because the charged-pion elastic electromagnetic form factor is well approximated by a monopole \cite{Wang:2025usl}.  It is a benchmark for the empirical expressions of ESIP.

It is worth recording the peak locations and heights of the curves in Fig.\,\ref{FpiKDF}:
\begin{equation}
\begin{array}{l|ccccc}
\zeta_2  & x {\mathpzc q}^{\rm sym} & x {\mathpzc q}^{\rm sf} & x {\mathpzc u}^\pi & x {\mathpzc u}^K & x \bar {\mathpzc s}^K \\ \hline
x{\rm -location} & 0.34 & 0.33 & 0.33(2) & 0.30(2) & 0.36(2)\\
{\rm height} & 0.41 & 0.47 & 0.42(2) & 0.42(2) & 0.47(2) \\
\end{array}.
\end{equation}

%
%


\smallskip

\noindent\textbf{DF Moments from Lattice-regularised  QCD}\,---\,%
An approach to quantisation of gauge field theories using a discrete lattice formulation in Euclidean spacetime was introduced in Ref.\,\cite{Wilson:1974sk}.  Since then, the scheme has been developed for use as a quantitative approach to solving QCD.
In that context, as highlighted elsewhere \cite[Sec.\,8]{Roberts:2021nhw}, methods have been proposed during the past decade, or so, that give access to the $x$-dependence of DFs via the numerical simulation of lattice-formulated QCD \cite{Ji:2014gla, Radyushkin:2017cyf, Ma:2017pxb, Sufian:2020vzb}.
However, today, these approaches deliver results that remain subject to substantial uncertainties: progress and challenges are described, \emph{e.g}., in Ref.\,\cite{Lin:2025hka}.
%
%

An alternative of longer standing is the direct calculation of DF Mellin moments \cite{Martinelli:1987si}.
This approach uses the operator product expansion \cite[OPE]{Wilson:1969zs} to map the bilocal operator associated with a given hadron's parton DFs into a sum of local operator contributions, each of which could potentially be calculated using lQCD.
It provides access to the following DF Mellin moments ($n\in \mathbb N_0$):
{\allowdisplaybreaks
\begin{subequations}
\label{MellinLattice}
\begin{align}
\langle x^{2n}\rangle_{{\mathpzc q}_V^H}^\zeta & =
\int_0^1 dx \, x^{2n} [ {\mathpzc q}^H(x;\zeta) -  \bar {\mathpzc q}^H(x;\zeta)]\,,
\\
\langle x^{2n+1}\rangle_{{\mathpzc q}_\Sigma^H}^\zeta & =
\int_0^1 dx \, x^{2n+1} [ {\mathpzc q}^H(x;\zeta) + \bar {\mathpzc q}^H(x;\zeta)]\,,
\\
\langle x^{2n+1}\rangle_{{\mathpzc g}^H}^\zeta & =
\int_0^1 dx \, x^{2n+1} {\mathpzc g}^H(x;\zeta)  \,.
\end{align}
\end{subequations}
This means that even moments are associated with valence (nonsinglet) DFs
and odd moments are projections of singlet DFs.
Notably, whereas valence quark/antiquark moments maintain their identities under DGLAP evolution, singlet quark and glue moments mix: interactions entail quark+antiquark$\,\leftrightarrow\,$glue$\,\leftrightarrow\,$glue+glue.
It is clear from Eq.\,\eqref{MellinLattice} that neither odd moments of nonsinglet distributions nor even moments of singlet distributions are readily accessible using this technique.
So-called disconnected contributions to matrix elements must be estimated/calculated in order to enable the separation of valence and singlet moments.
These are self-connected quark loops, linked only by glue to the lines joining the incoming and outgoing hadron constituents.  Their estimation requires sophisticated algorithms and the results typically suffer from high statistical noise.
This presents one impediment to lQCD extractions of DF moments.
}

In fact, an array of other hurdles, attendant upon the discretisation of spacetime that is inherent in the lQCD approach, not least of which is the loss of true $O(4)$ invariance (Euclidean Poincar\'e invariance), restricted such moment calculations to $n\leq 3$ and stalled progress for more than thirty years.
Today, novel algorithms are being explored, which proponents hope can provide reliable access to higher moments.
For instance, it is argued that a gradient flow method \cite{Francis:2025pgf} is capable of delivering precise results for moments up to $n=6$ and beyond.  Whether such optimism is justified is, as yet, unclear.
The existing study \cite{Francis:2025pgf} used an unphysically large pion mass ($m_\pi^2 \sim 10 \times \,$experiment) and delivered $n=5, 6$ moments with central values in conflict with the physical constraint in Eq.\,\eqref{MomOrdering}.  Moreover, its utility for the kaon has not yet been explored. For these reasons, we do not discuss this approach further herein.

\begin{figure*}

\centerline{%
\includegraphics[clip, width=0.95\textwidth]{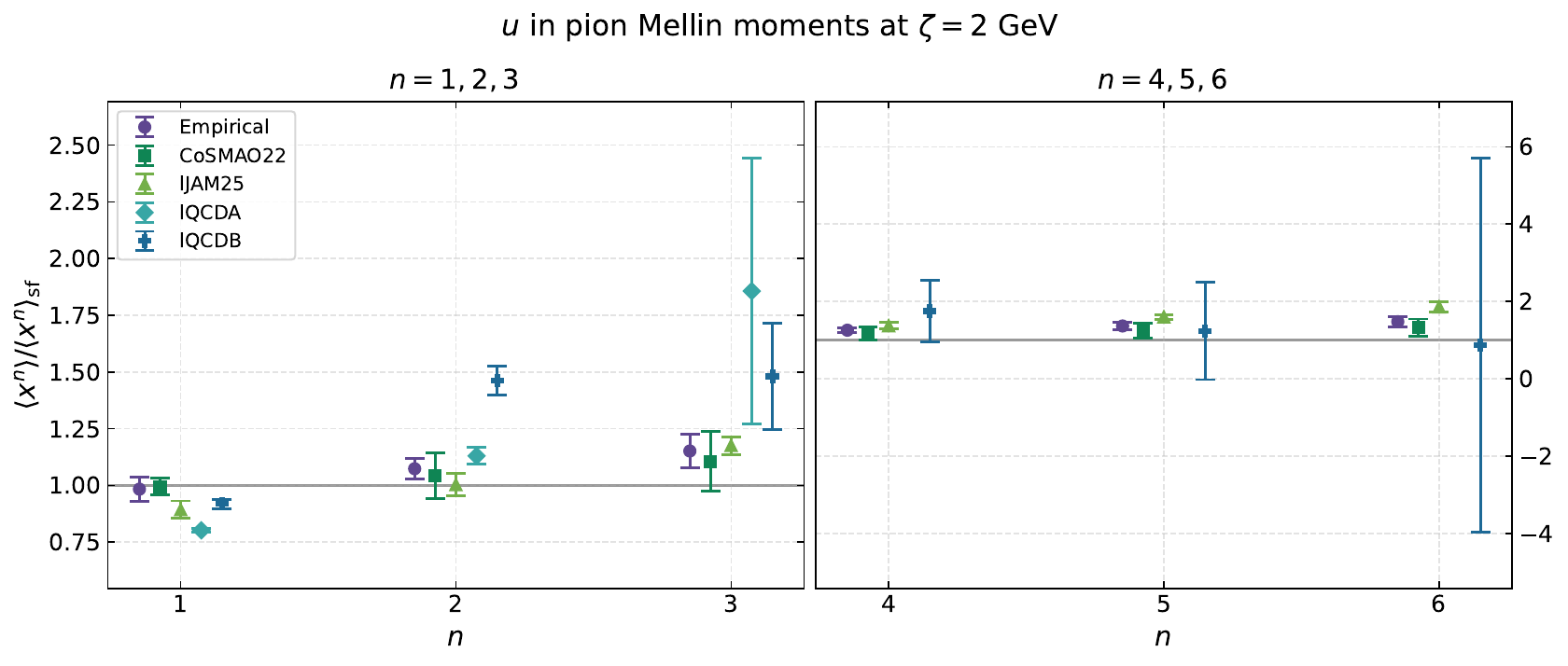}}

\vspace*{-2ex}

\caption{\label{F1tag}
In-pion $u$-quark Mellin moments.
Legend:
\cite[Empirical]{Xu:2024nzp};
\cite[CosMA022]{Cui:2020tdf};
\cite[lJAM25]{Barry:2025wjx, Alexandrou:2021mmi};
\cite[lQCDA]{Alexandrou:2024zvn, Alexandrou:2026nsl};
\cite[lQCDB]{Miller:2026hza}.
}
\end{figure*}

A typical lQCD simulation is defined by $3+n_f$ parameters:
lattice spacing, $a$;
lattice volume, which we will denote as $L^4$, although asymmetric lattices are used for some applications;
QCD coupling, $g$;
and the $n_f$ lQCD Lagrangian quark masses, where, typically, $n_f=4$ in studies that aim at realistic unified studies of $\pi$, $K$ observables.
In this formulation, $g=g(a)$ and all mass-dimensioned quantities are measured in units of $1/a$.
A physical scale is then set by computing a matrix element that delivers a value for some measured mass-dimensioned quantity.
These steps are repeated until the practitioner reaches a desired set of parameter values, which are as realistic as achievable within current computer capacity.
In connection with DFs, this set of parameters defines the lattice resolving scale, $\zeta_{\rm lQCD}$, which is then mapped onto a continuum scale using methods from perturbative QCD; hence, is only as accurate as the available form of that mapping.
Lattice results are commonly reported at $\zeta=\zeta_2$ using what is called the $\overline{\rm MS}$-scheme \cite{Pascual:1984zb}.
It should be noted that sea (nonvalence) quark contributions are known to be significant at this scale, with continuum theory and phenomenology agreeing that, \emph{e.g}., in the pion, sea quarks carry $10$\% or more of its light-front momentum at $\zeta_2$ \cite[Table~1]{Cui:2021mom}.

Naturally, numerical simulations deliver results with statistical errors.  Their size is readily determined.
More challenging is the assessment of lQCD systematic uncertainties,
%
of which there are many known sources, including
recovery of the continuum limit and symmetries via the extrapolations $a\to 0$ and $L\to \infty$;
extrapolation to physical quark current masses, which is often still needed;
and identification of clean/pure hadron-state signals (elimination of excited state contamination).
Furthermore and especially important for DFs, as noted above, errors associated with the estimation/calculation of disconnected contributions to DF-related matrix elements.

To illustrate the current status of lQCD results for pion and kaon DF moments, we consider two recent studies.
References~\cite{Alexandrou:2024zvn, Alexandrou:2026nsl} deliver moments up to $n=3$ using quark masses tuned to give the physical pion mass and standard OPE methods, with local operators chosen to avoid mixing between moments out to this order.
Only one ensemble of gauge-field configurations was employed, so continuum and infinite volume limit uncertainties are unknown.  Some other systematic uncertainties are estimated, but the budget is incomplete.
In the pion, for the $n=1$ moment, Refs.\,\cite{Alexandrou:2024zvn, Alexandrou:2026nsl} find the ratio of singlet-to-nonsinglet to be $1.11(14)$, confirming that disconnected (sea) contributions are substantial and that their neglect must introduce a significant uncertainty in any estimate of the first moment.
For $n=3$, Refs.\,\cite{Alexandrou:2024zvn, Alexandrou:2026nsl} assume the disconnected contributions are negligible.
CSMs predict them to be a $2$\% effect for $n=3$ \cite{Lu:2022cjx}; so, the assumption introduces an additional (albeit smaller) systematic error.

\begin{figure*}
\begin{tabular}{c}
\includegraphics[clip, width=0.95\textwidth]{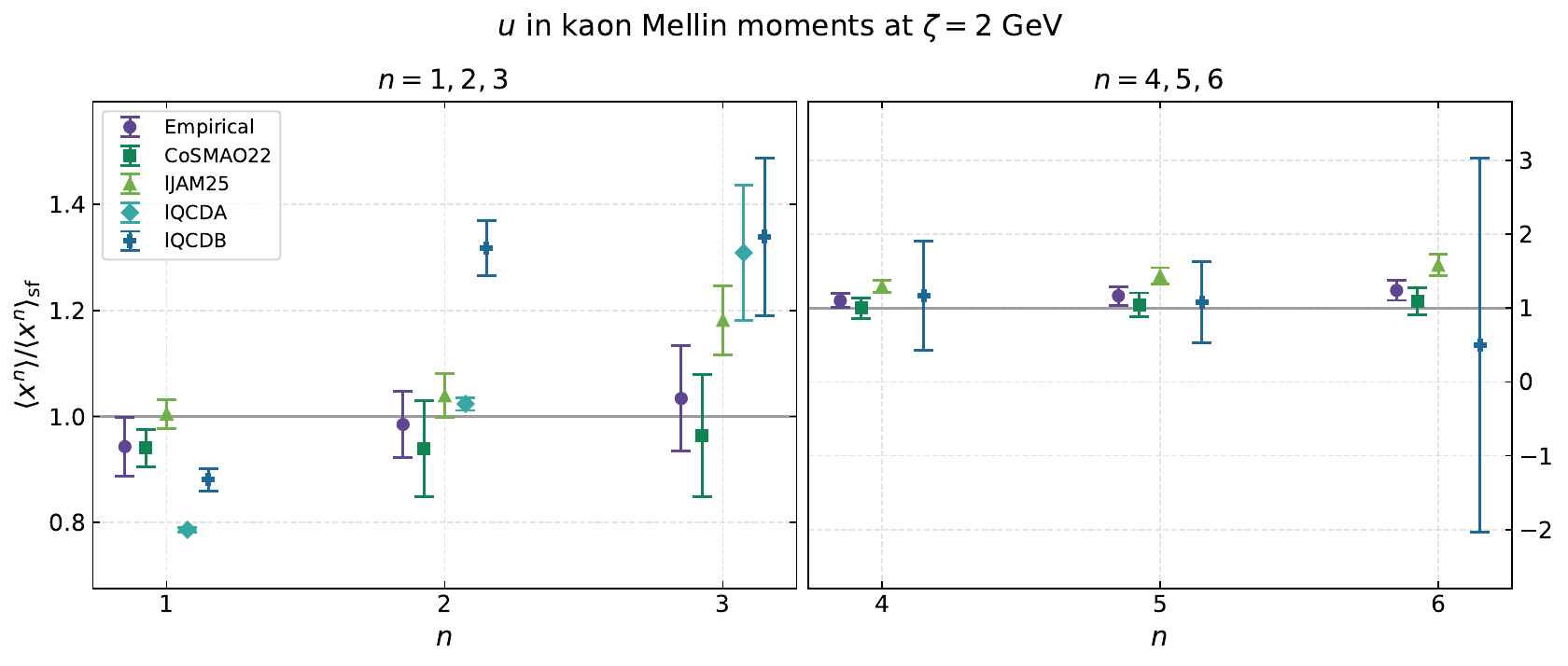} \\
\includegraphics[clip, width=0.95\textwidth]{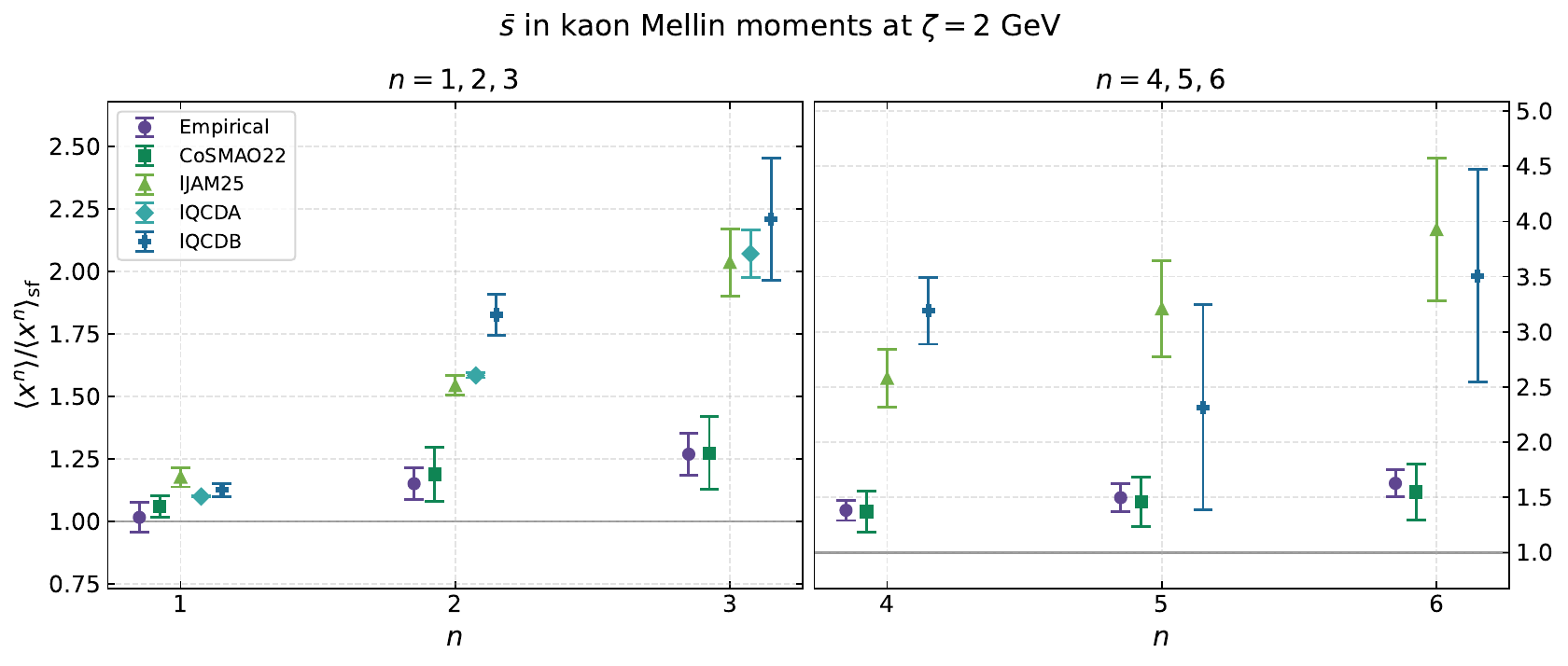}
\end{tabular}
\vspace*{-2ex}

\caption{\label{F4tag}
In-kaon $u$ and $\bar s$-quark Mellin moments.
Legend:
\cite[Empirical]{Xu:2024nzp};
\cite[CosMA022]{Cui:2020tdf};
\cite[lJAM25]{Barry:2025wjx, Alexandrou:2021mmi};
\cite[lQCDA]{Alexandrou:2024zvn, Alexandrou:2026nsl};
\cite[lQCDB]{Miller:2026hza}.
}
\end{figure*}

In contrast, Ref.\,\cite{Miller:2026hza} uses Ioffe-time pseudo-distributions (ITDs) \cite{Radyushkin:2017cyf} to relate boosted hadron operators to the Mellin moments of their DFs.  This approach enables one to circumvent local lattice operator mixing and so, in principle, reach $m>3$.
On the other hand, it faces potentially severe limitations in connection with, \emph{inter alia},
hadron-boost momentum resolution,
quadratic-divergence nonlocal operator mixing,
and statistical uncertainties connected with the need for large target hadron momentum.
These and related issues, like fitting strategies for DF moment reconstruction from ITD lattice outputs, lead to errors on moments that grow with increasing $n$.
Another source of uncertainty is contained in the need to use perturbative QCD truncations in relating lattice output to continuum values of DF moments.
Reference~\cite{Miller:2026hza} reports results out to $n=6$ obtained using only one ensemble of gauge-field configurations at a pion mass for which $m_\pi^2 \sim 3.6 \times \,$experiment.
%
%
It is probable, therefore, that the result obtained for each moment is statistically limited and possibly attended by significant but magnitude-unknown systematic effects.
%
%
Furthermore, only connected contributions to operator matrix elements were considered.
%
%
The omission of disconnected diagrams means that objective extraction of $m\in \mathbb O$ valence-quark moments is problematic, an issue that is highlighted but nonetheless disregarded in Ref.\,\cite{Miller:2026hza}.

\smallskip

\noindent\textbf{Predicted DF Mellin Moments for Pions and Kaons}\,---\,%
In the comparisons presented in this section we plot a given method's $n^{\rm th}$ Mellin moment divided by that obtained from Eq.\,\eqref{sfDF} (no-ESIP) after AO evolution $\zeta_{\cal H} \to \zeta_2$.  Using Eq.\,\eqref{MellinLattice}, these moments are:
\begin{equation}
\label{sfmoments}
\begin{array}{l|ccccccc}
n  & 0 & 1 & 2 & 3 & 4 & 5 & 6 \\ \hline
\langle x^n \rangle_{{\mathpzc q}^{\rm sf}_V}^{\zeta_2}
& 1 & & 0.0903 & & 0.0223 & & 0.00798
\\
\langle x^n \rangle_{{\mathpzc q}^{\rm sf}_\Sigma}^{\zeta_2}
& & 0.274 & & 0.0425 & & 0.0129 &
\end{array}
\end{equation}
(\emph{N.B}. The zeroth moment is a test of baryon number conservation, which all acceptable analyses must obey.)
Only if true uncertainty (which may be larger than reported uncertainty) significant deviations from unity are seen for all six nontrivial moment ratios can the subject set of moments be objectively argued to provide evidence for a DF whose pointwise behaviour is different from the SF curve in Fig.\,\ref{FpiKDF}.



In Fig.\,\ref{F1tag}, we compare $n=1,\ldots,6$ Mellin moments of the in-pion $u$ quark DF as obtained using lQCD-assisted phenomenological fits \cite[lJAM25]{Barry:2025wjx, Alexandrou:2021mmi}; empirical information \cite{Xu:2024nzp}; CSMs \cite{Cui:2020tdf}, and lQCD \cite{Alexandrou:2024zvn, Alexandrou:2026nsl, Miller:2026hza}.
The empirical moments show a significant signal for dilation of the in-pion $u$ DF with respect to the no-ESIP result, reflecting the actual result in Ref.\,\cite{Xu:2024nzp}.
Such dilation is a ``smoking gun'' expression of EHM dynamics.
A signal for slightly less dilation is seen in the CSM moments.
Within mutual uncertainties, the lJAM25 moments are consistent with the empirical and CSM moments; hence, objectively, regarding these moments, there is little signal for any difference between the lJAM25 phenomenological fit and the CoSMAO curve in Fig.\,\ref{FpiKDF}.

\begin{figure*}
\begin{tabular}{ccc}
{\large\sf A} \hspace*{26em} & $\;$ & {\large \sf B} \hspace*{26em} \\[-4ex]
\includegraphics[clip, width=0.46\textwidth]{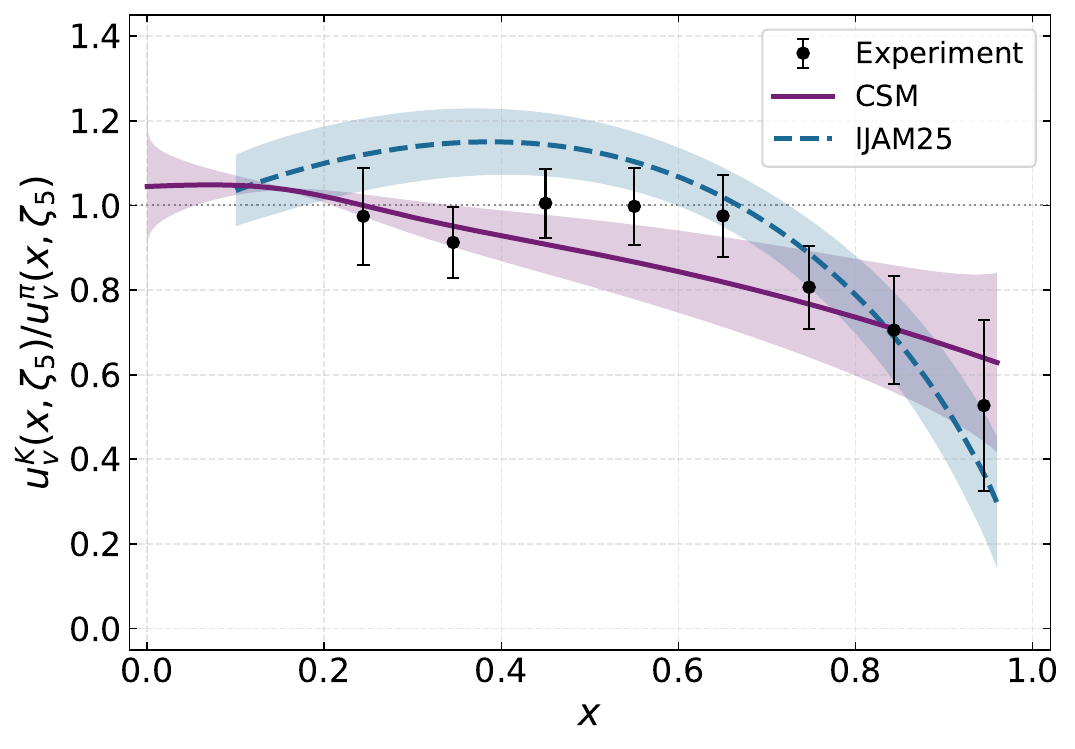} & $\;$ &
\includegraphics[clip, width=0.46\textwidth]{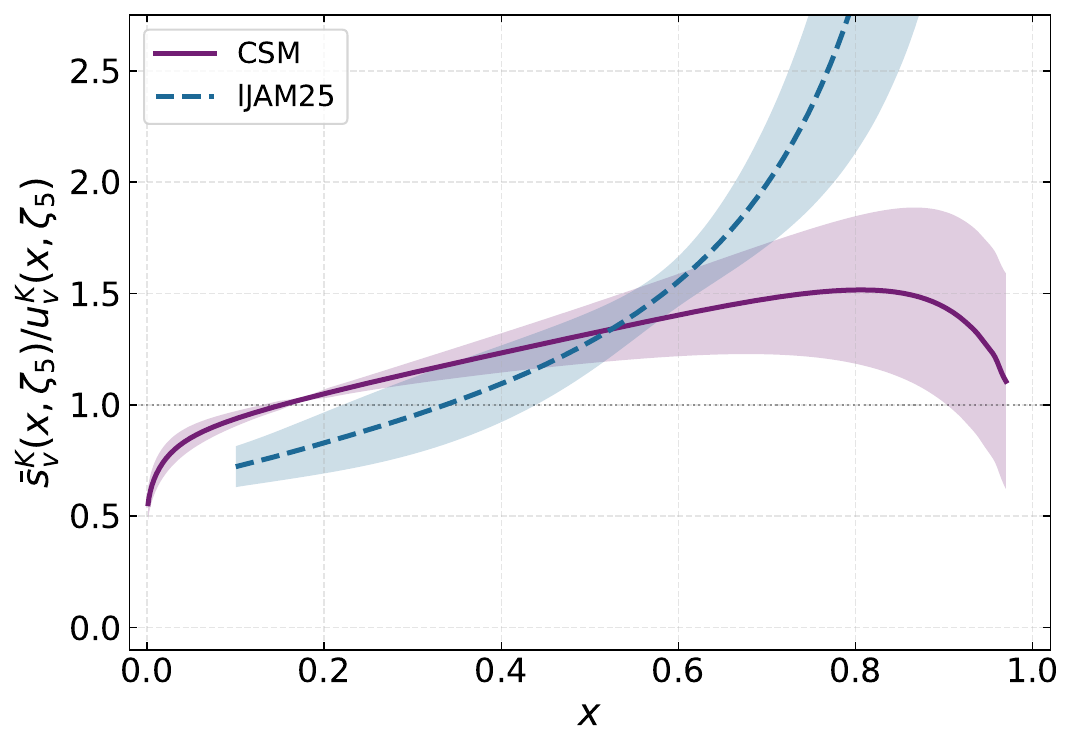}
\end{tabular}

\vspace*{-2ex}

\caption{\label{DYratio}
{\sf Panel A}.
Drell-Yan data on the kaon to pion structure function ratio \cite[NA3K]{Badier:1980jq} at resolving scale $\zeta_5 = 5.2\,$GeV.
Legend.
Solid purple curve and uncertainty band -- parameter-free CSM prediction for the ratio
$\bar{\mathpzc u}^K_V(x;\zeta_5)/\bar{\mathpzc u}^\pi_V(x;\zeta_5)$ \cite[CSM]{Cui:2020tdf};
dashed blue curve and associated band -- analogous curve computed from available phenomenological fits \cite[lJAM25]{Barry:2025wjx, Alexandrou:2021mmi}.
{\sf Panel B}.
$\bar{\mathpzc s}^K_V(x;\zeta_5)/{\mathpzc u}^K_V(x;\zeta_5)$:
CSM prediction -- solid purple curve and uncertainty band;
analogous curve from phenomenological fit -- dashed blue curve and associated band.
}
\end{figure*}

lQCDA delivers only 3 nontrivial moments, which paint an unusual picture.
The $n=1$ moment is unexpectedly low and its reported uncertainty appears too small.
On the other hand, the central value of the $n=3$ moment is remarkably big, albeit with a large thus far identified uncertainty.
The second moment is consistent with continuum theory and phenomenology.
Consequently, the results in Refs.\,\cite{Alexandrou:2024zvn, Alexandrou:2026nsl} do not enable any objective determinations of difference between the in-pion $u$-quark DF and the no-ESIP curve in Fig.\,\ref{FpiKDF}.
It follows that the pointwise DF reconstructions in Ref.\,\cite[Sec.\,V]{Alexandrou:2026nsl} and conclusions derived therefrom are subjective.

lQCDB delivers an $n$=1 moment that is, within mutual uncertainties, consistent with continuum theory and phenomenology.
However, the $n=2$ moment (purely valence) is then remarkably large in comparison.
%
%
The only way to reconcile these values is to suppose that the $u$-quark sea (and, so, the $d$-quark sea, too) momentum fraction is negative.
DF positivity precludes such an outcome.
%
%
%
Hence, it may be worth reviewing the internal consistency of the lQCDB moments.
%
%
Beyond this, the large thus far identified uncertainties (that on $n=6$ is 5.5-times larger than the central value) make subjective any conclusions about the pointwise behaviour of the in-pion $u$-quark DF \cite[Sec.\,IV\,D]{Miller:2026hza}.


The panels in Fig.\,\ref{F4tag} compare $n=1,\ldots,6$ Mellin moments of the in-$K^+$ $u$- and $\bar s$-quark DFs as obtained using lQCD-assisted phenomenological fits \cite[lJAM25]{Barry:2025wjx, Alexandrou:2021mmi}; empirical information \cite{Xu:2024nzp}; CSMs \cite{Cui:2020tdf}, and lQCD \cite{Alexandrou:2024zvn, Alexandrou:2026nsl, Miller:2026hza}.  (Owing to ${\cal G}$-parity symmetry, the analogous DFs in other kaons are identical.)
Referring back to Fig.\,\ref{FpiKDF}, one sees that the pattern of the empirical and CoSMAO moments reveals a modest shift of $x{\mathpzc u}^K(x;\zeta_2)$-support toward $x=0$ as compared with $x{\mathpzc u}^\pi(x;\zeta_2)$.

It is worth recalling Eq.\,\eqref{momfracs}, which can be restated as
$2 \langle x \rangle_{{\mathpzc u}^\pi}^{\zeta_{\cal H}}
=  \langle x \rangle_{{\mathpzc u}^K}^{\zeta_{\cal H}}
+ \langle x \rangle_{\bar {\mathpzc s}^K}^{\zeta_{\cal H}}$
because valence dof carry all properties of a given hadron at $\zeta_{\cal H}$.
Using kernels that are sensitive to quark current-mass effects, this identity need not be conserved under evolution.
However, with realistic kernels of this type \cite{Cui:2020tdf}, one finds that the right-hand side of Eq.\,\eqref{momfracs} is $\lesssim 2.5$\% larger than the left-hand side, with
$\langle x \rangle_{{\mathpzc u}^K}^{\zeta_2} \approx 0.96  \langle x \rangle_{{\mathpzc u}^\pi}^{\zeta_2}$,
$\langle x \rangle_{\bar {\mathpzc s}^K}^{\zeta_2} \approx 1.03 \langle x \rangle_{{\mathpzc u}^\pi}^{\zeta_2}$.
Plainly, therefore, the CSM $n=1$ moments in Figs.\,\ref{F1tag}, \ref{F4tag} are consistent with Eq.\,\eqref{momfracs}.  This is also true of the empirical $n=1$ moments.

On the other hand, the lJAM25 fit returns an in-kaon sum that is $19(1)$\% larger than the in-pion result.  This outcome stems from the fact that the lJAM25 $u$-in-pion $n=1$ moment is small (as discussed above) and both $u$ and $\bar s$ in-kaon $n=1$ moments are large.
Notably, for these fits, $\langle x \rangle^{\zeta_2}_{\bar{\mathpzc s}^K} = 1.33(1)\langle x \rangle^{\zeta_2}_{{\mathpzc u}^\pi}$, \emph{viz}.\ a moment value displacement $\approx 11$ times larger than the CSM prediction.

lQCDA reports an in-kaon $n=1$ sum that is $18(1)$\% larger than the in-pion result.  In this case, the imbalance arises because
$\langle x \rangle^{\zeta_2}_{{\mathpzc u}^K} \approx \langle x \rangle^{\zeta_2}_{{\mathpzc u}^\pi}$
whereas $\langle x \rangle^{\zeta_2}_{\bar{\mathpzc s}^K} = 1.38(1)\langle x \rangle^{\zeta_2}_{{\mathpzc u}^\pi}$.
Recall, too, that the lQCDA $u$-in-$\pi$ $n=1$ moment may be unnaturally low (see above).

Regarding the lQCDB results, the in-kaon sum is $9(1)$\% bigger than the in-pion result.  This imbalance is smaller than that produced by lJAM25 and lQCDA, but it is still significantly larger than the CSM prediction ($4$-times the displacement).
In this case, the cause is found in the fact that
$\langle x \rangle^{\zeta_2}_{\bar {\mathpzc s}^K} = 1.23(1)\langle x \rangle^{\zeta_2}_{{\mathpzc u}^\pi}$, whereas, in line with the CSM prediction,
$\langle x \rangle_{{\mathpzc u}^K}^{\zeta_2} \approx 0.96  \langle x \rangle_{{\mathpzc u}^\pi}^{\zeta_2}$.

Considering the upper panel in Fig.\,\ref{F4tag} ($u$-in-$K$) and taking all six nontrivial moments together, one sees that, within mutual uncertainties, the lJAM25 values are consistent with the empirical and CSM moments.  Thus, there is no significant signal for any difference between the lJAM25 phenomenological fit and the CoSMAO curve for $x{\mathpzc u}^K(x)$ in Fig.\,\ref{FpiKDF}B.

\begin{figure*}
\hspace*{-0.2em}\includegraphics[clip, width=1.0\textwidth]{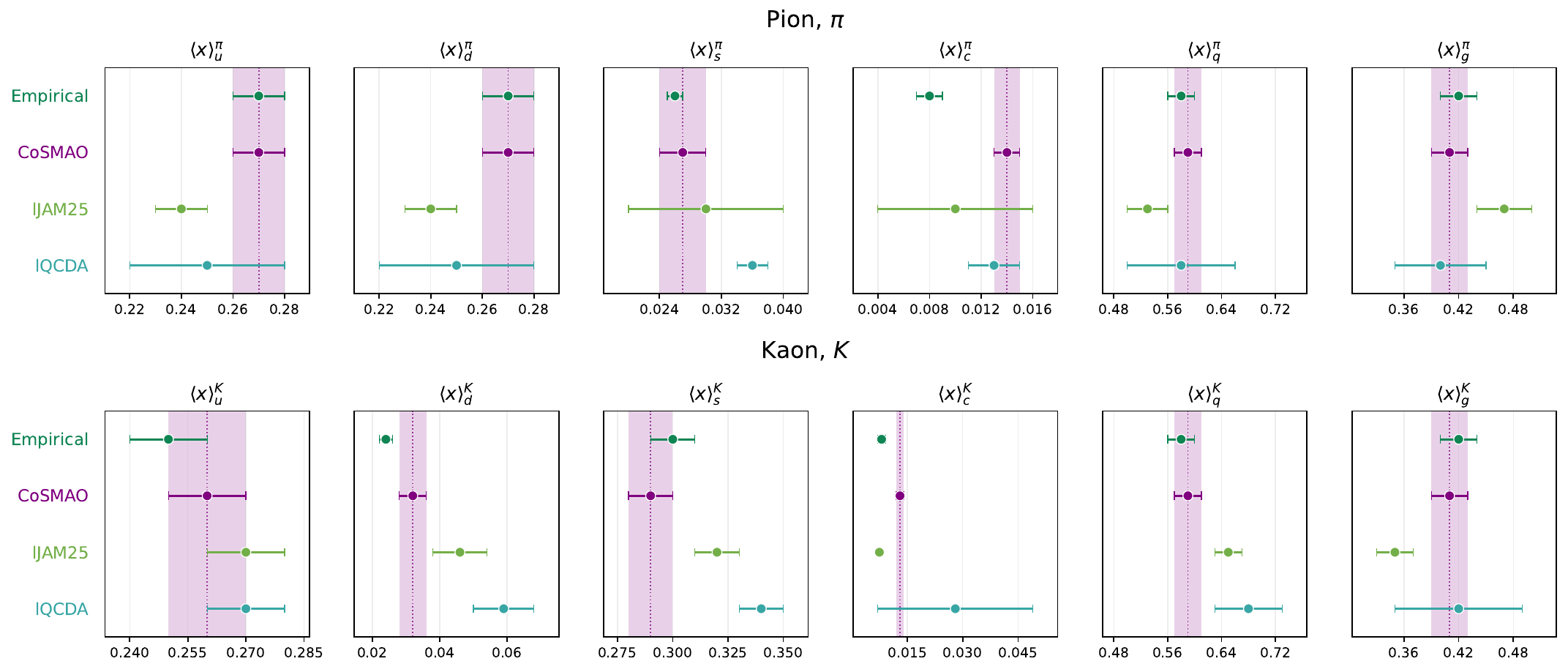}

\vspace*{-2ex}

\caption{\label{Fgluefractions}
Species-separated light-front momentum fractions for the pion (upper) and kaon (lower).
Legend:
\cite[Empirical]{Xu:2024nzp, Yin:2023dbw};
\cite[CoSMAO]{Cui:2020tdf};
\cite[lJAM25]{Barry:2025wjx, Alexandrou:2021mmi};
\cite[lQCDA]{Alexandrou:2024zvn}.
The purple bands mark the CoSMAO (CSM) prediction and associated uncertainty.
}
\end{figure*}

\begin{table*}[t]
\centering
\caption{Species separations of in-pion and in-kaon light-front momentum fractions.
Legend:
\cite[Empirical]{Xu:2024nzp, Yin:2023dbw};
\cite[CoSMAO]{Cui:2020tdf};
\cite[lJAM25]{Barry:2025wjx, Alexandrou:2021mmi};
\cite[lQCDA]{Alexandrou:2024zvn}.
CoSMAO and lJAM25 uncertainties are fully anticorrelated; so, in both cases, the uncertainty on the sum is zero (final column).
\label{gluefractions}}
\begin{tabular*}
{\hsize}
{
l|@{\extracolsep{0ptplus1fil}}
l@{\extracolsep{0ptplus1fil}}
l@{\extracolsep{0ptplus1fil}}
l@{\extracolsep{0ptplus1fil}}
l|@{\extracolsep{0ptplus1fil}}
l@{\extracolsep{0ptplus1fil}}
l|@{\extracolsep{0ptplus1fil}}
l@{\extracolsep{0ptplus1fil}}}\hline
$\pi\ $   & $\langle x\rangle_{\mathpzc u}\ $ & $\langle x\rangle_{\mathpzc d}\ $ & $\langle x\rangle_{\mathpzc s}\ $ & $\langle x\rangle_{\mathpzc c}\ $
& $\langle x\rangle_{\mathpzc q}\ $
& $\langle x\rangle_g\ $
& $\langle x\rangle_{\mathpzc q}+ \langle x\rangle_g\ $
\\\hline
empirical  & $0.27(1)\ $ & $0.27(1)\ $ & $0.026(1)\ $ & $0.008(1)\ $ & $0.58(2)\ $ & $0.42(2)\ $ & $1.0\ $\\
CoSMAO  & $0.27(1)\ $ & $0.27(1)\ $ & $0.027(3)\ $ & $0.014(1)\ $ & $0.59(2)\ $ & $0.41(2)\ $ & $1.0\ $ \\
lJAM25 &  $0.24(1)\ $ & $0.24(1)\ $ & $0.03(1)\ $ & $0.010(6)\ $ & $0.53(3)\ $ & $0.47(3)\ $ & $1.0\ $\\
lQCDA & $0.25(3)\ $ & $0.25(3)\ $ & $0.036(2)\ $ & $ 0.013(2)\ $ & $0.58(8)\ $ & $0.40(5)\ $ & $0.98(9)\ $ \\
\hline
$K\ $   & $\langle x\rangle_{\mathpzc u}\ $ & $\langle x\rangle_{\mathpzc d}\ $ & $\langle x\rangle_{\bar {\mathpzc s}}\ $ & $\langle x\rangle_{\mathpzc c}\ $
& $\langle x\rangle_{\mathpzc q}\ $
& $\langle x\rangle_g\ $
& $\langle x\rangle_{\mathpzc q}+ \langle x\rangle_g\ $
\\\hline
empirical  & $0.25(1)\ $ & $0.024(2)\ $ & $0.30(1)\ $ & $0.008(1)\ $ & $0.58(2)\ $ & $0.42(2)\ $ & $1.0\ $\\
CoSMAO  & $0.26(1)\ $ & $0.032(4)\ $ & $0.29(1)\ $ & $0.013(1)\ $ & $0.59(2)\ $ & $0.41(2)\ $& $1.0\ $ \\
lJAM25 & $0.27(1)\ $ & $0.046(8)\ $ & $0.32(1)\ $ & $0.0074(3)\ $ & $0.65(2)\ $ & $0.35(2)\ $& $1.0\ $ \\
lQCDA & $0.27(1)\ $ & $0.059(9)\ $ & $0.34(1)\ $ & $ 0.028(21)\ $ & $0.68(5)\ $ & $0.42(7)\ $ & $1.13(11)\ $ \\
\hline
\end{tabular*}
\end{table*}

Once again, the three moments available from the lQCDA computation, with $n=1$ low and possessing a small reported uncertainty but $n=3$ high with a large identified uncertainty, paint an unclear picture.
Thus the pointwise DF reconstructions in Ref.\,\cite[Sec.\,V]{Alexandrou:2026nsl} and conclusions derived therefrom may be seen as subjective.

Regarding the lQCDB moments in Fig.\,\ref{F4tag}\,-\,upper panel, the $n$=1 moment is, within mutual uncertainties, consistent with continuum theory, as we have already remarked, but inconsistent with phenomenology.
%
However, again, the purely valence $n=2$ moment is remarkably large in comparison; so indicating
that the in-kaon $u$-quark sea momentum fraction is negative, a possibility that is precluded by DF positivity.
%
Hence, it may also be worth reviewing the internal consistency of the lQCDB $u$-in-$K$ moments.
Beyond this, the large reported uncertainties on $n=3,4,5,6$ make subjective any conclusions about the pointwise behaviour of the in-$K$ $u$-quark DF \cite[Sec.\,IV\,D]{Miller:2026hza}, as was also the case with the attempted reconstruction of the $u$-in-$\pi$ DF.

Turning to Fig.\,\ref{F4tag}\,-\,lower panel, one sees that all analyses predict that the $\bar s$-in-$K$ DF has more support at large $x$ than ${\mathpzc q}^{\rm sf}(x;\zeta_{2})$,
\emph{viz}.\ owing to Higgs boson couplings into QCD, the $\bar s$ quark number density is displaced toward $x=1$.
However, this is where the consensus ends.
The lJAM25 fit and lQCDA and lQCDB computations are consistent with an $x\to 1$ relocation of support in $x\bar {\mathpzc s}^K(x;\zeta_2)$ which is far larger than that predicted by CSMs \cite{Cui:2020tdf} and found in the AO evolution analysis of available empirical information \cite{Xu:2024nzp}.
It is worth noting that the agreement between lJAM25 and lQCDA moments is not entirely coincidental: the Ref.\,\cite{Alexandrou:2021mmi} computation was used both to constrain the Ref.\,\cite{Barry:2025wjx} fits and as input for the lQCDA analysis \cite{Alexandrou:2024zvn, Alexandrou:2026nsl}.

\smallskip

\noindent\textbf{Ratios Involving Kaon Valence Quark DFs}\,---\,%
A Drell-Yan measurement of the kaon to pion structure function ratio is reported in Ref.\,\cite[NA3K]{Badier:1980jq} at resolving scale $\zeta_5 = 5.2\,$GeV.
Interpreted as the ratio $\bar{\mathpzc u}^K_V(x;\zeta_5)/\bar{\mathpzc u}^\pi_V(x;\zeta_5)$, the eight reported points represent \emph{all} that is directly known about in-kaon parton structure.
The data are plotted in Fig.\,\ref{DYratio}A.

This figure also depicts the CSM prediction from Ref.\,\cite{Cui:2020tdf} and the analogous curve drawn in Ref.\,\cite[lJAM25]{Barry:2025wjx} on its domain of reported validity.
Notably, the CSM prediction approaches a constant value as $x\to 1$ because all in-pion and in-kaon valence dof DFs exhibit the same large-$x$ behaviour, \emph{viz}.\ Eq.\,\eqref{DFpower}.
The lJAM25 fits do not respect this constraint: the ratio vanishes as $x\to 1$; \emph{i.e}., the fit suggests that, in the kaon, $u$ quarks play no role on the far valence domain; instead, they contribute only to the sea.
Evidently, as observed elsewhere \cite{Cui:2020tdf}, far more and more precise data are necessary before one can distinguish between these two curves; and data relevant to ${\mathpzc s}^K(x;\zeta_5)$, ${\mathpzc u}^K(x;\zeta_5)$, ${\mathpzc u}^\pi(x;\zeta_5)$ separately would be of even greater value.

Figure~\ref{DYratio}B compares the CSM prediction for $\bar{\mathpzc s}^K_V(x;\zeta_5)/{\mathpzc u}^K_V(x;\zeta_5)$ with the analogous curve built from lJAM25 fits on the their domain of reported validity.
The CSM prediction expresses a modest pointwise difference between the in-kaon valence DFs.
Moreover, again in line with Eq.\,\eqref{DFpower}, this ratio approaches a constant value as $x\to 1$.
The lJAM25 fit result is inconsistent with this constraint and delivers a huge discrepancy between in-kaon valence dof DFs on $x\gtrsim 0.7$.
Overall, if the phenomenological lJAM25 fits are a veracious representation of the SM pointwise behaviour of $\bar{\mathpzc s}^K_V(x)$, then Higgs boson couplings into QCD have a very large impact on kaon internal structure.

\smallskip

\noindent\textbf{Glue in Pions and Kaons}\,---\,%
The distributions of glue in pions and kaons also remain uncertain.
For instance, using currently available QCD-connected reaction models, analyses of pion-induced $J/\psi$ production suggest \cite{Chang:2020rdy} that the (hard) glue DFs inferred via early phenomenological fits \cite{Sutton:1991ay, Gluck:1999xe} are required to explain the data, with more recent fits and predictions, possessing softer large-$x$ behaviour, being disfavoured.
(In our view, this may again point to inadequacies in existing reaction models.)
Furthermore, working with early DF fits \cite{Gluck:1999xe}, Ref.\,\cite{Chen:2016sno} developed a picture of pion and kaon structure in which the glue content of the pion is far greater than that in the kaon.
The past few years have seen quite some attention to this issue from both theory and phenomenology; see, \emph{e.g}., Refs.\,\cite{Chang:2021utv, Han:2024yzj, NieMiera:2025inn, Kaur:2025gyr}.

\begin{figure}
\centerline{\includegraphics[clip, width=0.46\textwidth]{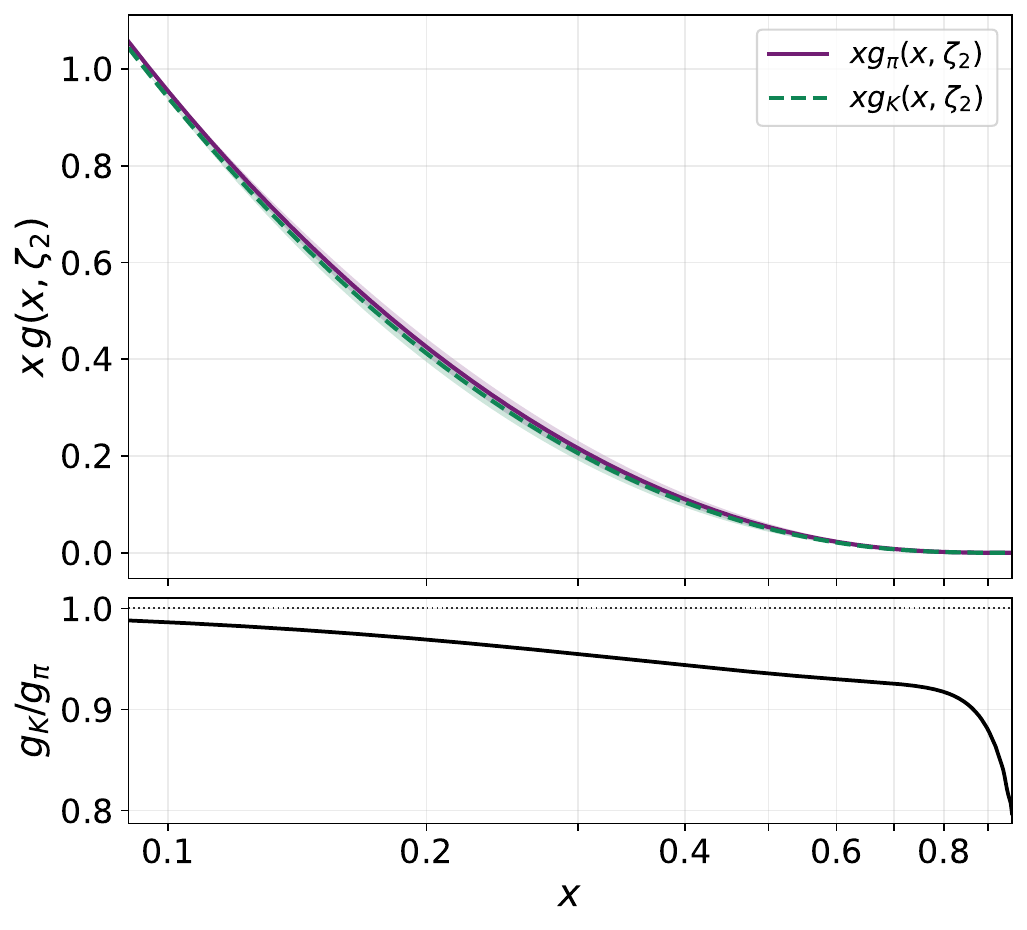}}

\vspace*{-2ex}

\caption{\label{F7tag}
$x$-weighted gluon DFs in the pion and kaon at $\zeta=\zeta_2$.
At large-$x$, the DFs behave as $(1-x)^{\beta_g}$, with $\beta_g \gtrsim 3$.
The lower panel depicts the ratio of the two curves in the upper panel.
(Curves are from CoSMAO22 at LHAPDF.)
}
\end{figure}

The Ref.\,\cite{Chen:2016sno} analysis was revisited in Ref.\,\cite{Cui:2020tdf}.
Exploiting modern CSM developments, including implementation of AO evolution \cite{Yin:2023dbw}, thereby making the study entirely self-sufficient, Ref.\,\cite{Cui:2020tdf} delivered a markedly different outcome.
Namely, that the in-pion and in-kaon glue momentum fractions are identical; see Fig.\,\ref{Fgluefractions} and Table~\ref{gluefractions}.
(In this figure and table, the empirical and CSM results include the Pauli blocking effect on evolution discussed elsewhere \cite{Yin:2023dbw}.)
Notwithstanding, there are small differences between the in-pion and in-kaon glue DF $x$-dependences; see Fig.\,\ref{F7tag}.

Considering the in-pion momentum fractions (Fig.\,\ref{Fgluefractions}\,-\,upper and Table~\ref{gluefractions}), further illustrating the discussion above, lJAM25 light-quark momentum fractions are significantly less than the CSM prediction.  Owing to momentum conservation, the lJAM25 glue fraction is therefore inflated.
Within their large identified uncertainties, the lQCDA results are consistent with the CSM predictions.

Turning to the kaon (Fig.\,\ref{Fgluefractions}\,-\,lower and Table~\ref{gluefractions}), and again confirming the discussion above, the lJAM25 fits locate more momentum with the $\bar s$ quark than do the predictions made by continuum analyses.
Consequently, the lJAM25 net-quark momentum fraction is larger and glue fraction smaller.
Notwithstanding, lJAM25 and continuum analyses are in agreement on the fact that glue momentum fractions in the pion and kaon are commensurate in magnitude.
The lQCDA results are only barely consistent with the momentum sum rule.
Given that the glue fraction is the same as the CSM prediction, then the excess may arise because the calculation locates an excess of momentum with the $\bar s$ quark, \emph{e.g}.,
if one divides $\langle x\rangle_{\bar {\mathpzc s}^K}$ by $\langle x\rangle_{{\mathpzc q}^K}+ \langle x\rangle_{g^K}$, then the result is $0.30(6)$, consistent with the continuum predictions.

\smallskip

\noindent\textbf{Summary and Perspective}\,---\,%
Continuum Schwinger function methods (CSMs) deliver parameter-free predictions for all pion and kaon parton distribution functions and relate their pointwise behaviour directly to the strong interaction dynamics responsible for the emergence of hadron mass and structure.  The bridge to quantum chromodynamics (QCD) is strong because the truncations employed in calculating the relevant matrix elements are mathematically sound approximations and have been validated in numerous other hadron observable calculations.
The CSM analyses predict that
(\emph{a}) the pointwise behaviour of pion and kaon DFs is principally determined by emergent hadron mass (EHM) dynamics;
(\emph{b}) the differences between them, which owe to Higgs boson couplings into QCD, are modest;
and (\emph{c}) those differences are expressed in a 20\% shift in support of light-quark valence degree-of-freedom DFs toward light-front momentum fraction $x=0$ and a compensating shift in $\bar s$-quark support toward $x=1$.

In order to confirm the CSM predictions, one needs more and more precise data, which can be related to pion and kaon DFs.  High luminosity facilities in operation or anticipated promise to deliver such data.
Meanwhile, phenomenological fits work with the limited available data set.

The results obtained from such fits are very sensitive to the assumed form of the reaction model that should, in principle, connect data to QCD predictions.  Today, whilst one choice of the reaction model delivers results in agreement with CSM predictions, another choice does not.  The latter choice also delivers a  Higgs-boson-induced modulation of EHM expressions in kaon DFs that is much larger than CSM predictions lead one to expect.  This uncertainty means that CSM predictions remain untested by data.  Indeed and more generally, unless robust, QCD-connected reaction models can be developed, then no amount of precise data will enable QCD predictions for in-hadron DFs to be validated.

Developments in the numerical simulation of lattice-regularised QCD (lQCD) continue, with practitioners expressing optimism that it may (soon) be possible to extract or infer precise information about the pointwise behaviour of in-hadron parton DFs.  Given the challenges faced in constructing a realistic reaction model for use in fitting data, the entire hadroparticle physics community is unified in desiring to see this hope realised.  Today, however, the goal remains distant; and whilst qualitative agreement does exist between CSM predictions and available lQCD results, quantitative agreement is lacking.  Indeed, as we illustrated herein, it is often the case that lQCD results are as much in agreement with CSM predictions as with phenomenological fits, so that both sets of results, which disagree with each other, may nevertheless claim support from extant lQCD studies.

Only the future, then, will see any fully substantiated, objective understanding of the structure of Nature's most fundamental (would be) Nambu-Goldstone bosons and the realisation of its promise to reveal secrets of EHM.
So, the quest continues.

\smallskip

\noindent\textit{Acknowledgments}\,---\,%
%
Work supported by:
National Natural Science Foundation of China, grant no.\ 12135007;
and Spanish Ministry of Science and Innovation (MICINN), grant no.\ PID2022-140440NB-C22).

\providecommand{\newblock}{}

\end{document}